\documentclass[12pt,preprint]{aastex63}
\usepackage{graphicx}
\usepackage[]{xcolor}

\received{...}
\revised{...}
\accepted{...}
\submitjournal{ApJ}

\begin{document}

\title{Did the Event Horizon Telescope Detect the Base of the Sub-Milliarsecond Tubular Jet in M\,87?}

\correspondingauthor{Brian Punsly}
\email{brian.punsly@cox.net}

\author{Brian Punsly}
\affiliation{1415 Granvia Altamira, Palos Verdes Estates CA, USA 90274}
\affiliation{ICRANet, Piazza della Repubblica 10 Pescara 65100, Italy}
\affiliation{ICRA, Physics Department, University La Sapienza, Roma, Italy}
\author{Sina Chen}
\affiliation{Physics Department, Technion Haifa, 32000, Israel}

\begin{abstract}
A high sensitivity, 7mm Very Long Baseline Array image of M\,87 was previously analyzed in order to estimate the bulk flow jet velocity between 0.4 and 0.65 mas from the point of origin using the asymmetry between the well-characterized double-ridged counter-jet (unique to this image) and the double ridged jet. We use this same image to estimate the cross-sectional area of this tubular stream. The velocity, acceleration, cross-sectional area and flux density along this stream determines a unique, perfect magnetohydrodynamic jet solution that satisfies, conservation of energy, angular momentum and mass (a monotonic conversion of Poynting flux to kinetic energy flux along the jet). The solution is protonic and magnetically dominated. The bilateral jet transports $\approx 1.2\times10^{-4} M_{\odot}/\rm{yr}$ and $\approx 1.1\times10^{42}$ erg/sec, placing strong constraints on the central engine. A Keplerian disk source that also produces the Event Horizon Telescope (EHT) annulus of emission can supply the energy and mass if the vertical magnetic field at the equator is $\sim 1-3.5$ G (depending on location). A Parker spiral magnetic field, characteristic of a wind or jet, is consistent with the observed EHT polarization pattern. Even though there is no image of the jet connecting with the annulus, it is argued that these circumstances are not coincidental and the polarized portion of the EHT emission is mainly jet emission in the top layers of the disk that is diluted by emission from an underlying turbulent disk. This is a contributing factor to the relatively low polarization levels that were detected.
\end{abstract}


\section{Introduction}

The galaxy M\,87 harbours the nearest powerful radio source ($\approx 16.8$ Mpc distant) making it a prime laboratory to study the physics of jet launching. This is a motivation for the continued development of the Event Horizon Telescope (EHT) \citep{aki23}. Despite the intense astronomical interest, basic physical parameters of the visible sub-lt-yr scale (tubular) jet are unknown. For example, the contemporaneous power of the tubular jet within years to decades of the EHT observation, its fraction of the total jet power and mass flux are unknown. These are important constraints for explaining the observed EHT emission, the currently detected annulus (EHT annulus, hereafter) as well as future detected emission. To this end, we develop a detailed physical picture of the tubular jet derived from the extraordinarily sensitive,  high dynamic range, 2013, 43 GHz Very Long Baseline Array (VLBA) observation \citep{wal18}. The bulk flow velocity field, 0.38 lt-yr to 0.61 lt-yr from the central black hole, in 2013 (4 years before the EHT polarization measurements were made), was estimated using special relativistic kinematics \citep{pun21}. We continue to exploit this image by extracting the dimensions of the partially resolved tubular jet. Using these dimensions, the perfect magnetohydrodynamic (MHD) conservation laws of mass, energy, angular momentum and magnetic flux, and the velocity, $v(z)$ ($z$ is the displacement along the jet axis), we find a best fit solution to the conservation laws that reproduces the jet surface brightness.
\par The jet power and mass flux in the sub-lt-yr scale jet have a natural origin in the nearby accretion disk. We assess the implications of these circumstances in the context of the polarization pattern and intensity of the EHT annulus in 2017 \citep{aki19,aki21}. We take an independent approach from the EHT collaboration given our newly attained vantage point. Instead of using a library of MHD numerical simulations to guide our interpretation of the EHT data, we use the kinematics of the adjacent tubular jet to bias the interpretation.
\par The first discussion describes the intensity cross-sections and how they supplement previous data to determine the jet physical properties. In Section 3, we cull through a myriad of possible solutions to find the most accurate solution to the MHD conservation laws. Sections 4 and 5 consider possible sources of the jet in the context of the observed properties of the EHT annulus. We adopt, a black hole mass, $M_{bh} \approx 6.6\times 10^{9}M_{\odot}$, corresponding to a geometrized mass $M\approx 9.74\times 10^{14}$cm, and a line of sight (LOS) to the jet axis of $18^{\circ}$ \citep{geb11}.

\section{Measurement of the High Sensitivity Image}
A high dynamic range image was obtained of M\,87 with 43 GHz VLBA on January 12, 2013 \citep{wal18}. Previously published images were restored with a mild super-resolution of 70\% of the major axis yielding a convolving beam 0.21 x 0.16 mas at PA = $0^{\circ}$ \citep{wal18,pun21}. This is the only image of M\,87 that shows a well-defined $\sim 1$mas long double-ridged counter-jet. The image is recreated again in Figure 1, for details see \citep{wal18,pun21}. The intensity asymmetry of the double ridged jet and counter-jet and the nuclear activity during the ejection times were considered to estimate the velocity field in \citet{pun21}, column (8) of our Table 1. Figure 1 shows the cross-sections at the locations chosen for analysis in \citep{pun21}. The over-sampling, every 0.05 mas, (beam width is 0.16 mas) was intentionally implemented to smooth any large fluctuations in dissipation induced by the environment and nuclear variability.
\begin{figure}
\begin{center}
\includegraphics[width=73mm, angle=0]{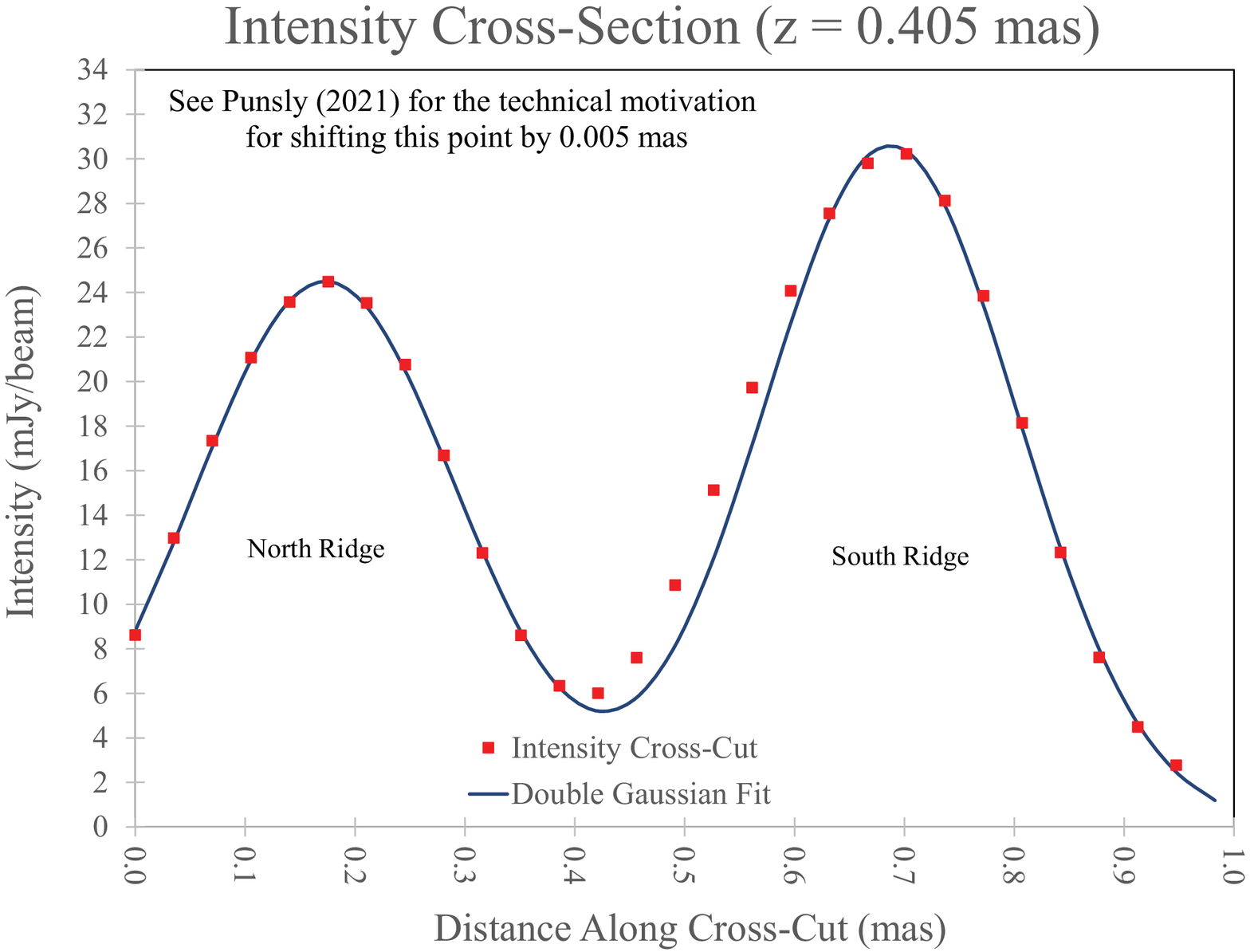}
\includegraphics[width=73mm, angle=0]{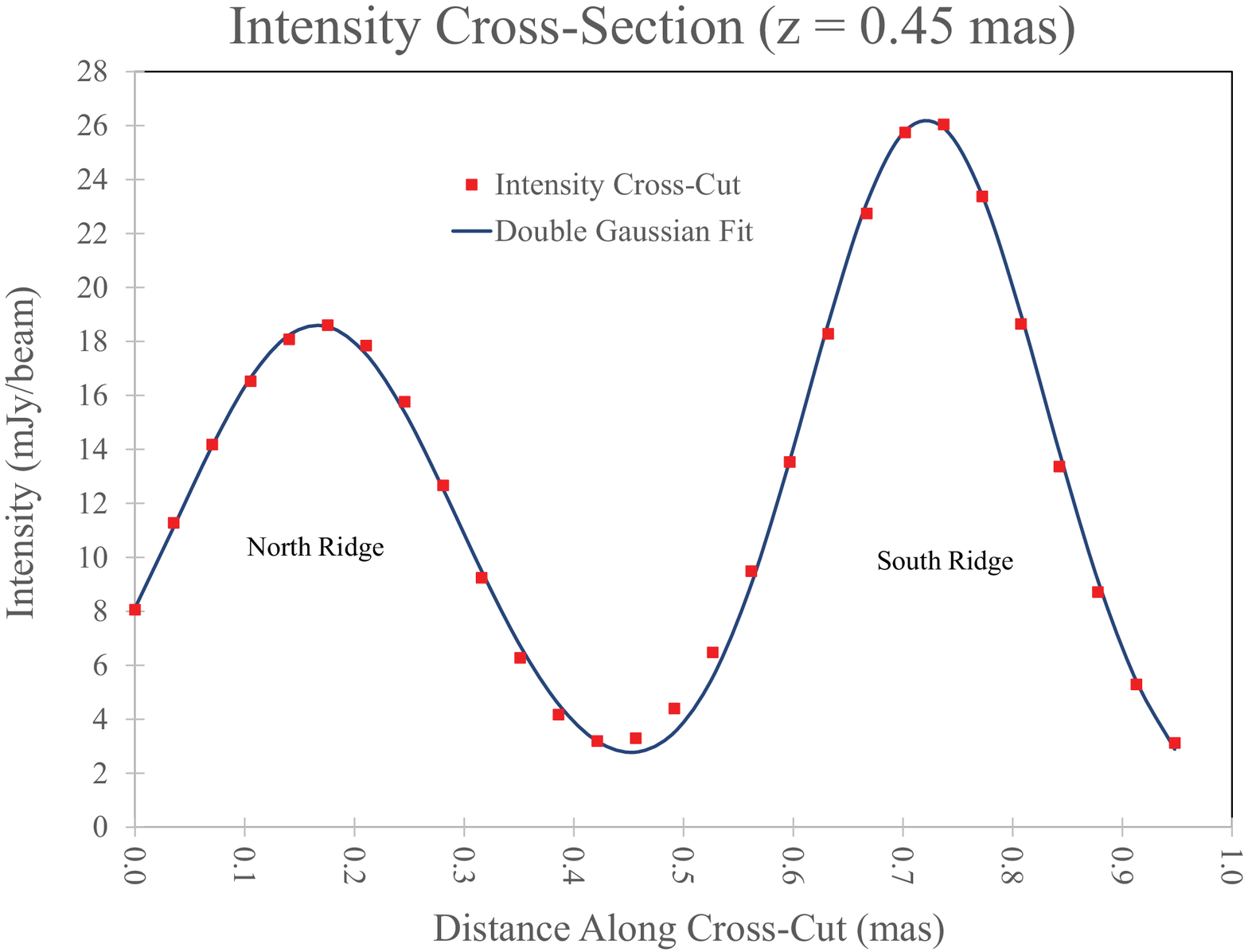}
\includegraphics[width=73mm, angle=0]{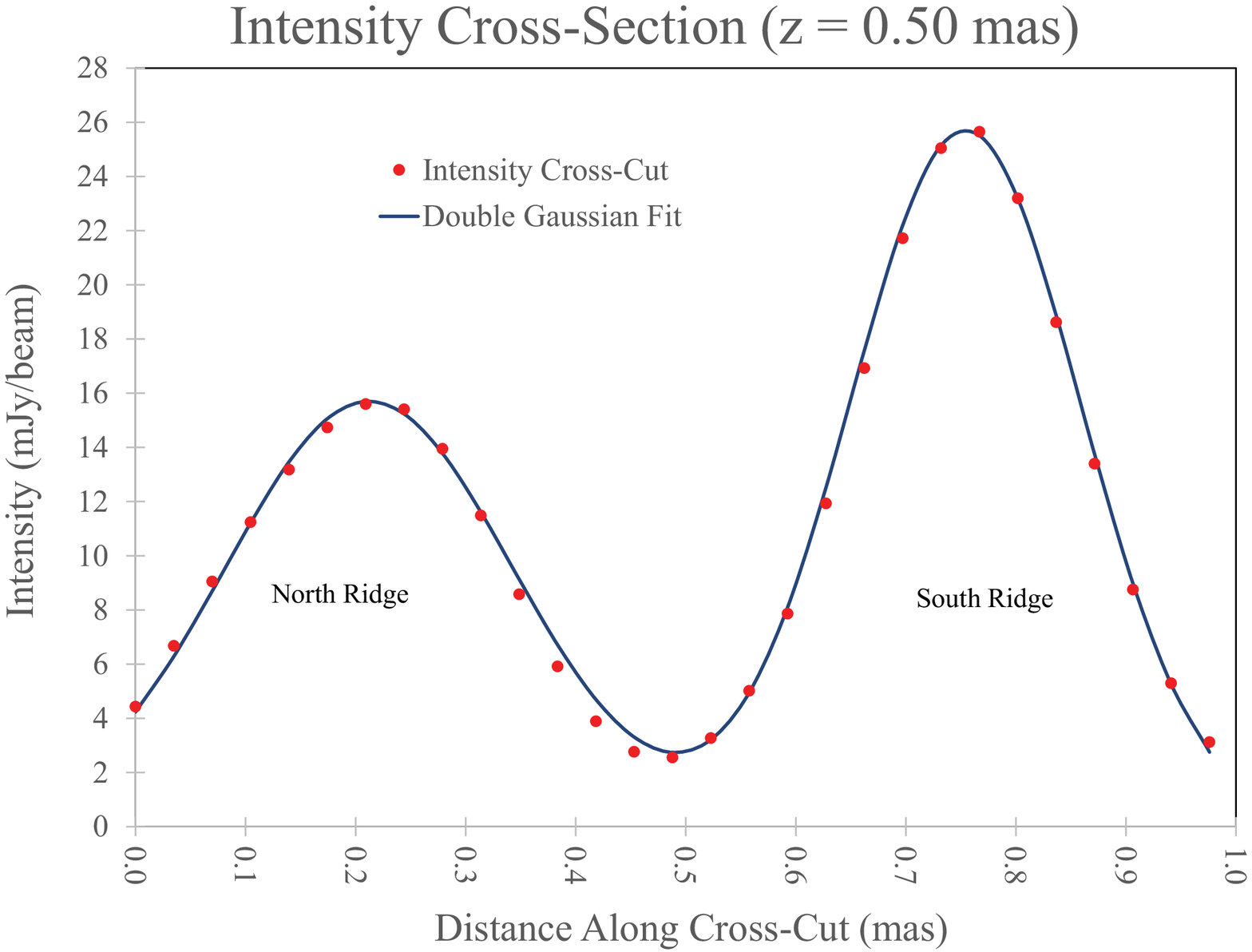}
\includegraphics[width=72mm, angle=0]{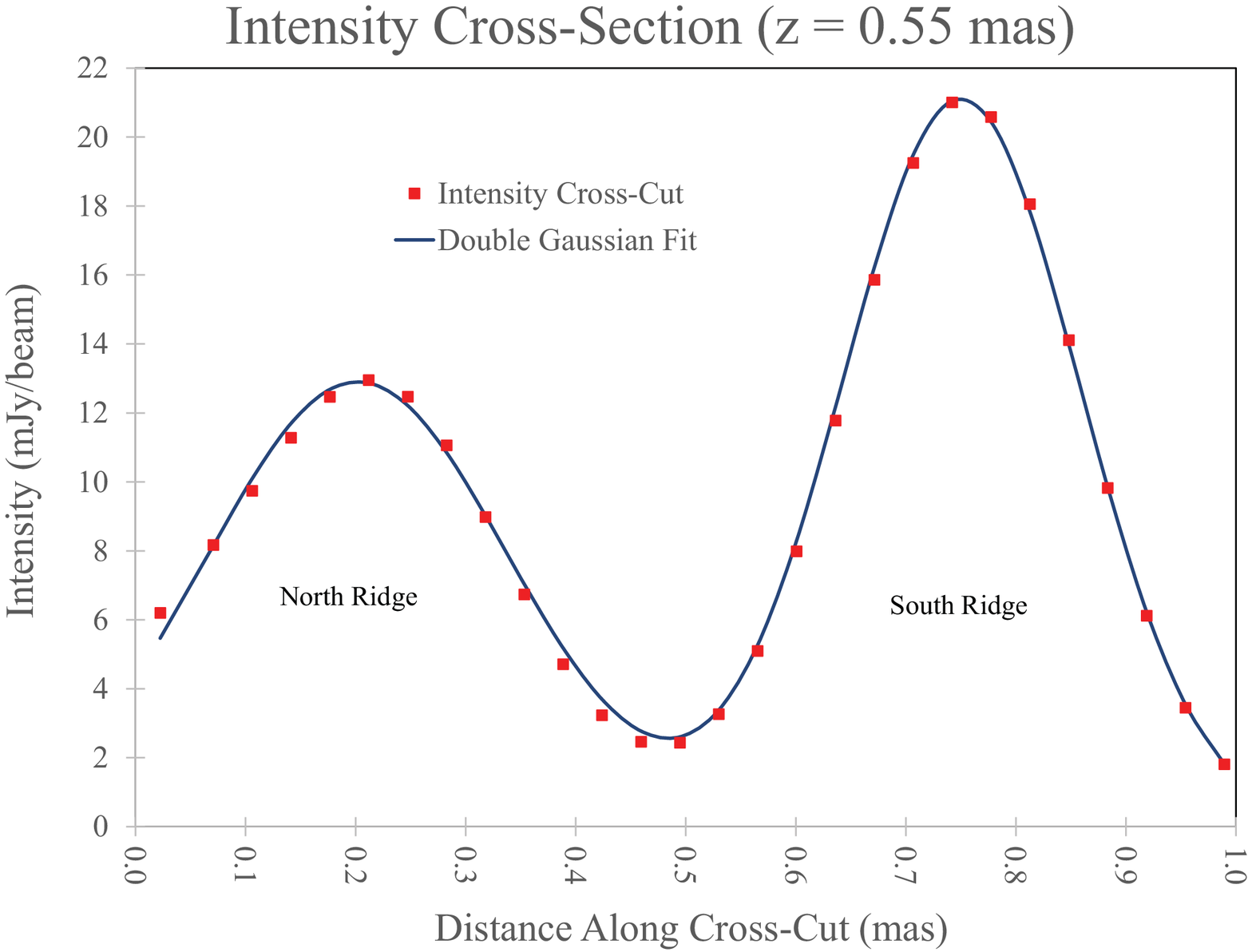}
\includegraphics[width=73mm, angle=0]{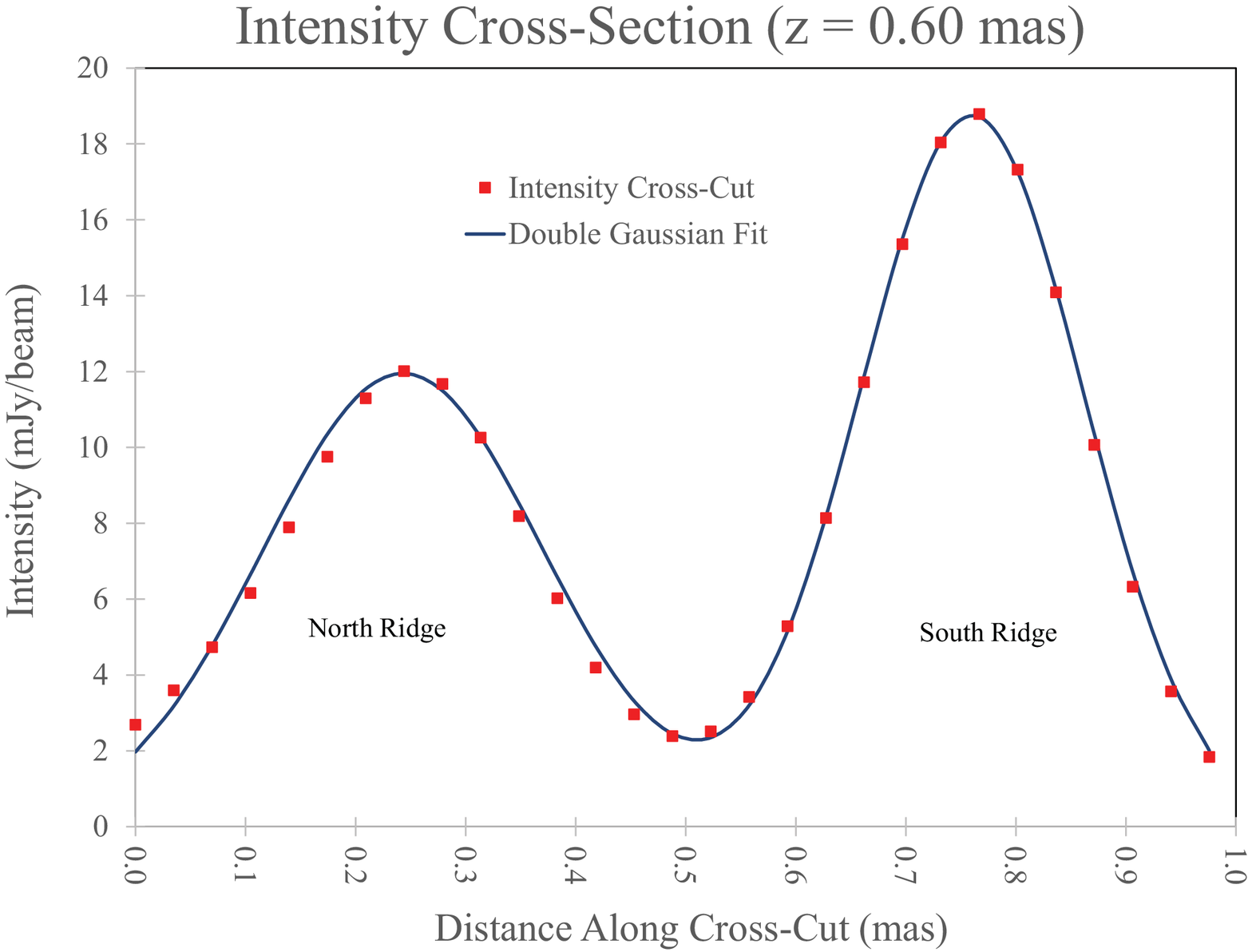}
\includegraphics[width=73mm, angle=0]{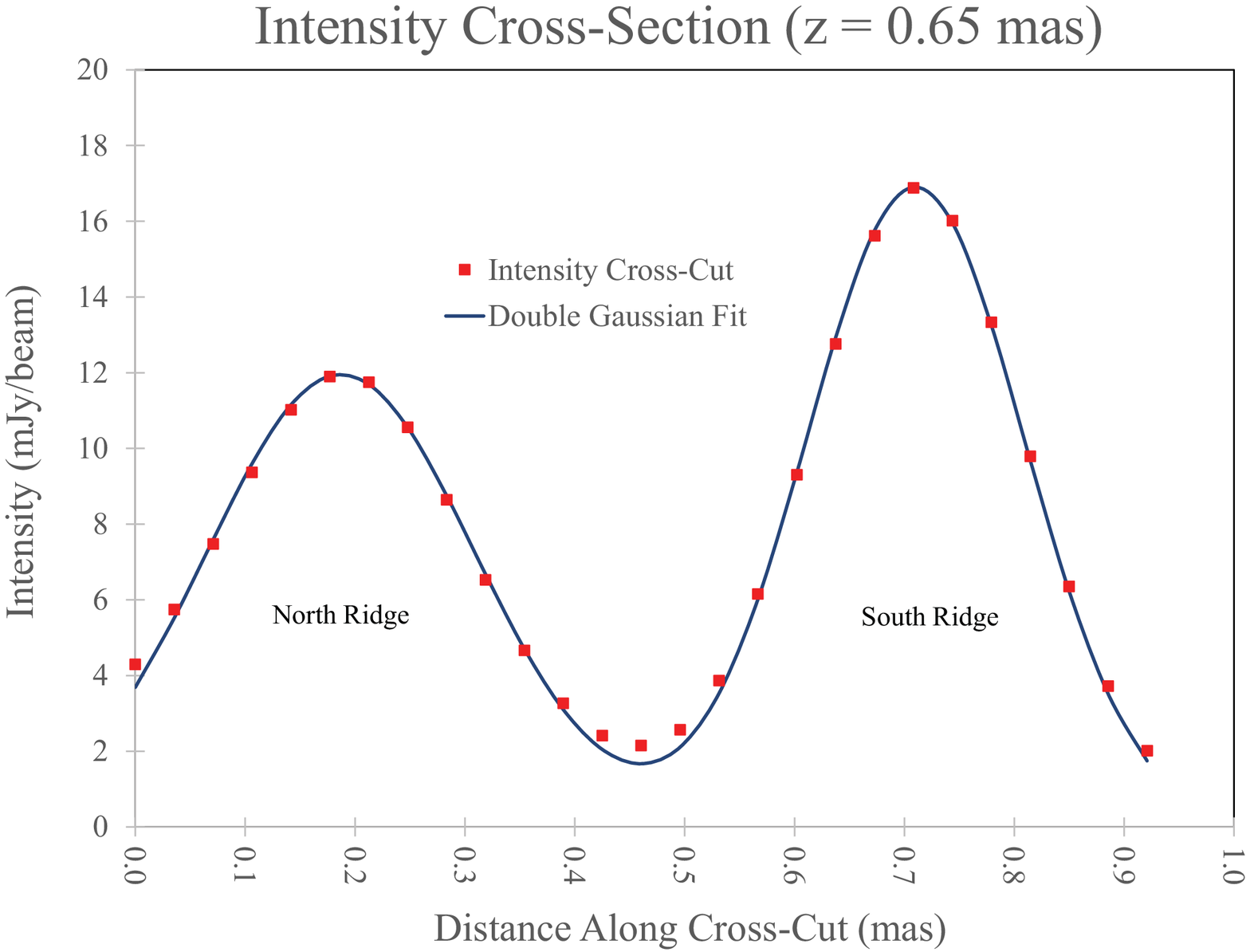}
\includegraphics[width=52mm, angle=0]{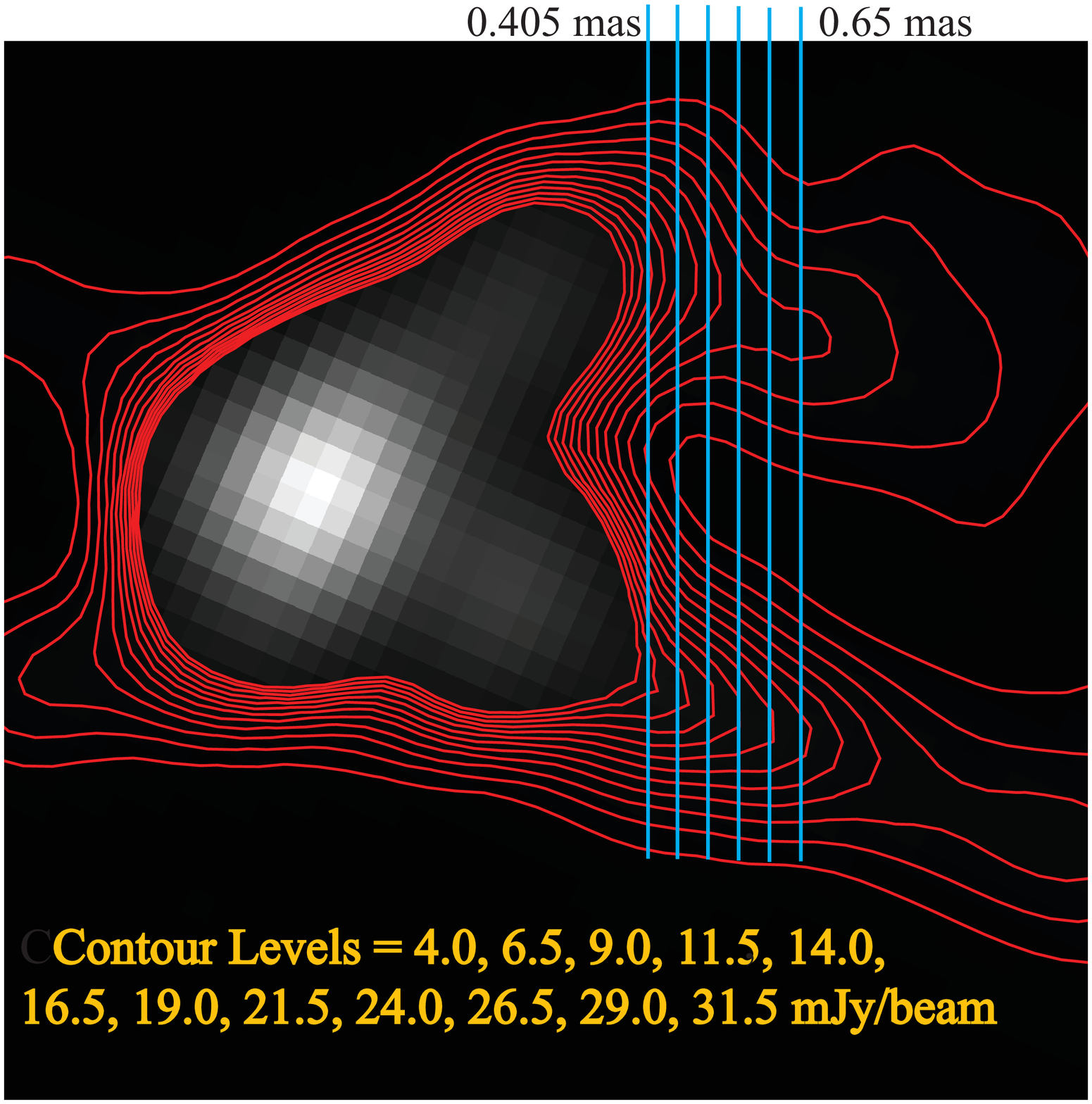}
\includegraphics[width=76mm, angle=0]{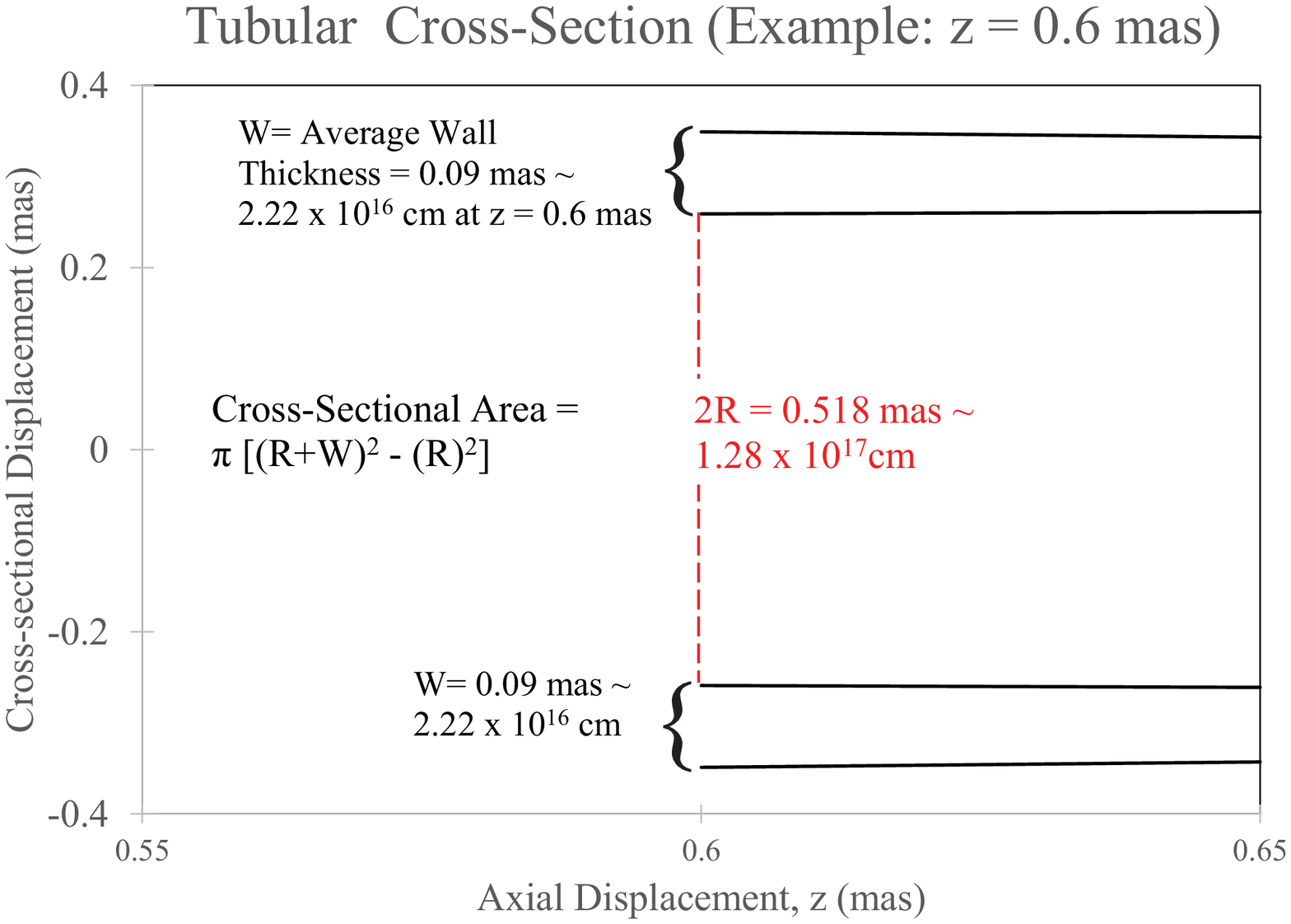}
\caption{The first three rows are the intensity cross-sections considered in Table 1. The cross-cuts are fit with two Gaussian sources. The location of the cross-cuts are indicated in the image in the lower left. An example of the cross-sectional area calculation appears in the lower right.}
\end{center}
\end{figure}
Two Gaussian (line) sources along the cross-cuts are convolved the restoring beam to obtain the fits in Figure 1. The full width at half maximum (FWHM) of each source is given in columns (2) and (3) of Table 1. The uncertainty is the minimum FWHM of a Gaussian that can be resolved based on the rms noise of 0.067 mJy/beam \citep{wal18,lob05}. A conical tube would have the brightest emission near the longest LOS through the tubular plasma, the inner edge of the tube \citep{wal18}. We confirmed this with numerous simplified cylindrical tubular sources of intensity and found the observed peak intensity (after convolution with the restoring beam) was within $\sim 0.015$mas of the inner edge of the tube ($<$ the column (5) uncertainty). Thus, we approximate the average thickness of the tubular walls as,
\begin{equation}
W=\rm{average\, thickness}\approx \rm{average}\,\rm{FWHM(wall)}/2 \equiv (FWHM(north)+ FWHM(south))/4\;,
\end{equation}
listed in column (4). The distance between the peaks is in column (5). Column (6) is the relative wall thickness that was previously estimated as 0.25 \citep{wal18}. The cross-sectional area of the tubular jet (see bottom right Figure 1) is estimated in column (7). Column (9) are the flux densities from Table 1 of \citep{pun21} in 0.16 mas wide strips (matched to the restoring beam width) obtained by adding CLEAN components.
\begin{table}
\begin{center}
 \caption{Physical Properties of the Tubular Jet}
\tiny{\begin{tabular}{cccccccccc}
 \hline
 (1)   & (2) & (3) & (4)  & (5) & (6) &  (7)& (8) & (9) \\
 $z$  & FWHM of & FWHM of &W, average & $2R$\tablenotemark{\tiny{a}} & $\frac{W}{R+W}$ & Cross-sectional & Axial & Flux Density \\
 & Gaussian source & Gaussian source &thickness of  & separation && area of & velocity & of 0.16 mas wide \\
 & north ridge & south ridge &tubular wall & of ridges & &tubular jet  &  & cross-section \\
  (mas)   & (mas) & (mas) & (mas)  & (mas) & &  ($\rm{cm}^{2}$)& (c) & (mJy) \\
 \hline
 0.405 & $0.203\pm 0.027$ & $0.186\pm 0.025$ & $0.097\pm 0.009$ & $0.514\pm 0.040$ & 0.27 & $1.14\pm0.11 \times 10^{34}$ & 0.270 & $66.4 \pm 6.7$ \\
0.450 & $0.229 \pm 0.032$ & $0.158 \pm 0.027$ & $0.097\pm 0.010$ & $0.555\pm 0.040$ & 0.26 & $1.21\pm0.13 \times 10^{34}$& 0.290 & $52.2 \pm 5.3$\\
 0.500 & $0.235 \pm 0.034$ & $0.146 \pm 0.027$ & $0.095\pm 0.011$  & $0.542\pm 0.040$ & 0.26 & $1.16 \pm 0.13\times 10^{34}$ & 0.315 & $44.6 \pm 4.6$ \\
 0.550 & $0.253\pm 0.038$ & $0.158\pm 0.030$ & $0.103\pm 0.012$ & $0.548\pm 0.040$ & 0.27 & $1.28\pm 0.15 \times 10^{34}$ & 0.340 & $33.5 \pm 3.5$\\
0.600 &$0.226\pm 0.040$ & $0.134\pm 0.032$ & $0.090\pm 0.013$  & $0.518\pm 0.040$ & 0.26 & $1.05\pm 0.15 \times 10^{34}$ & 0.360 & $32.2 \pm 3.4$ \\
0.650 & $0.206\pm 0.040$ & $0.121\pm 0.033$ & $0.082\pm 0.013$  & $0.523\pm 0.040$ & 0.24 & $9.46\pm 1.50 \times 10^{33}$ & 0.380 & $33.2 \pm 3.4$
\end{tabular}}
\end{center}
\tablenotetext{a}{uncertainty from \citet{lis09}}
\end{table}

\begin{figure*}[htp!]
\begin{center}
\includegraphics[width=85mm, angle=0]{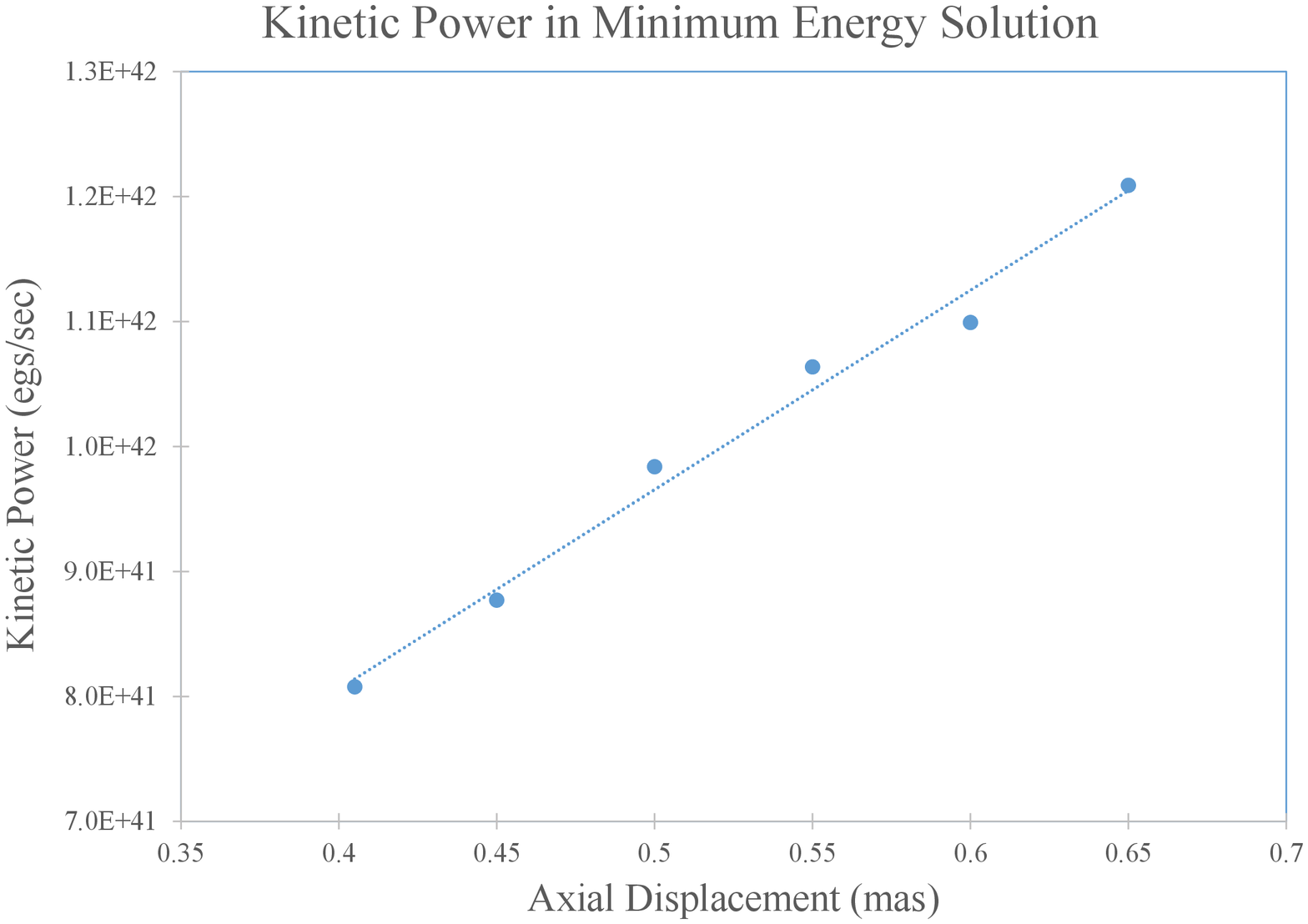}
\includegraphics[width=85mm, angle=0]{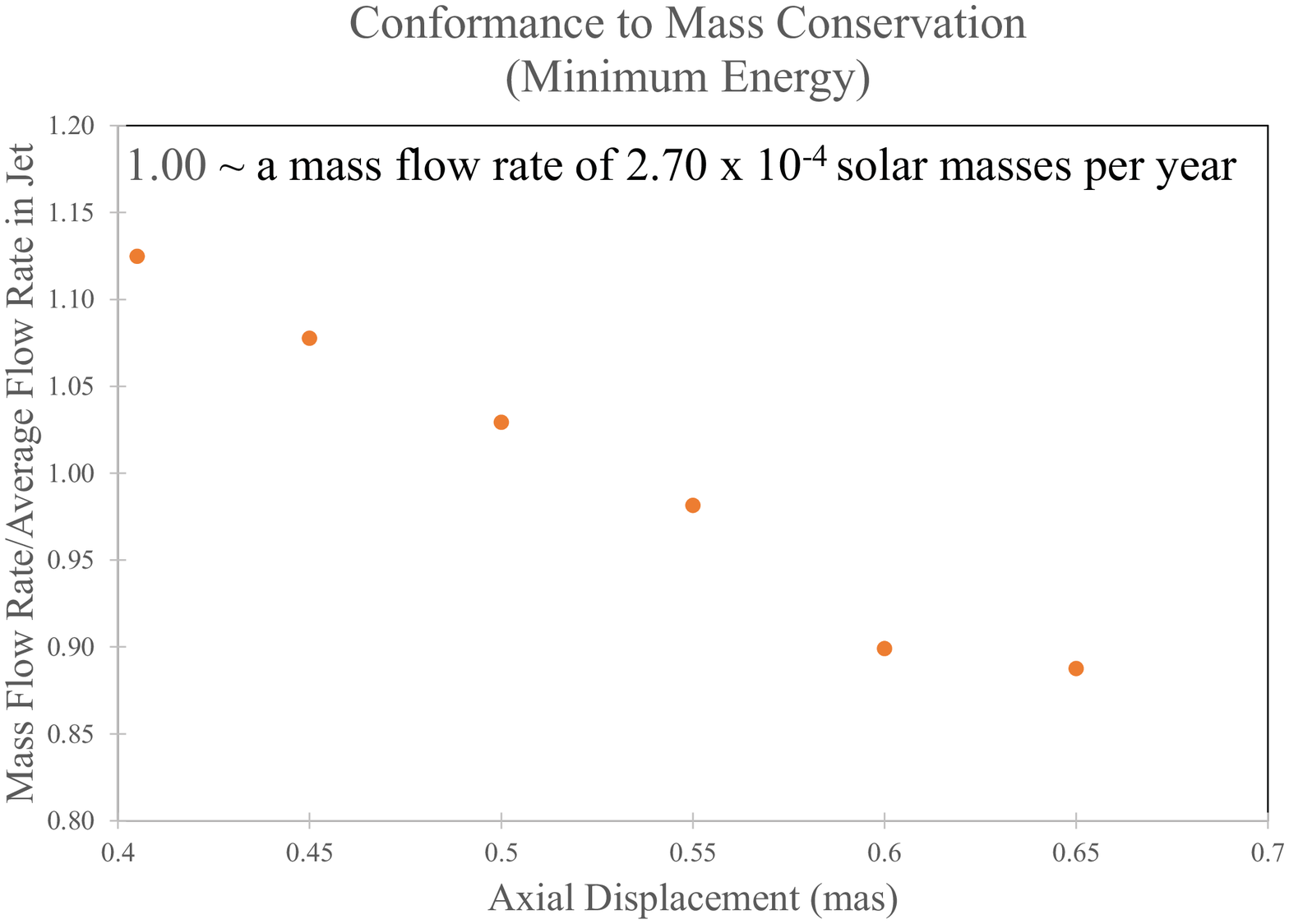}
\includegraphics[width=85mm, angle=0]{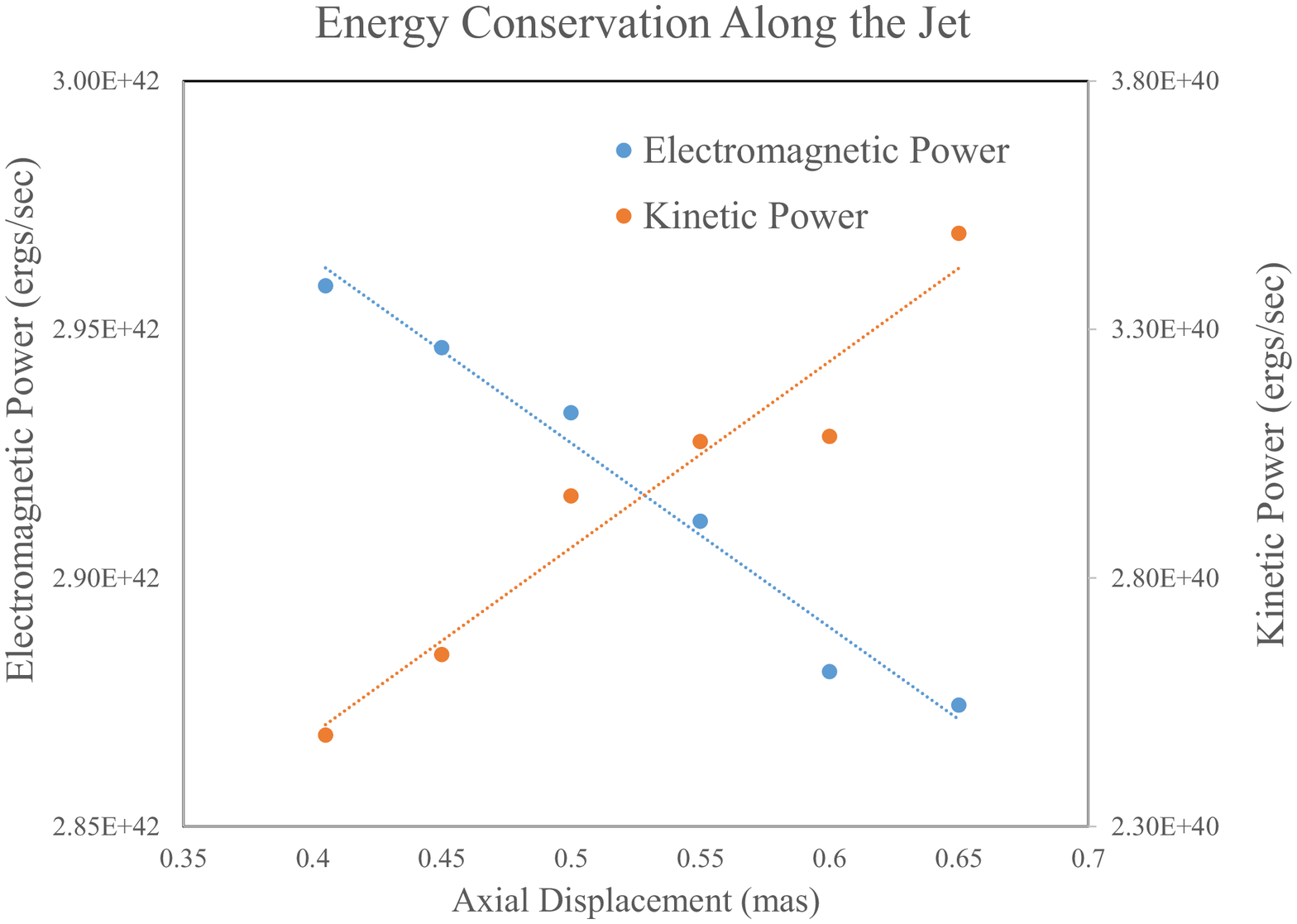}
\includegraphics[width=85mm, angle=0]{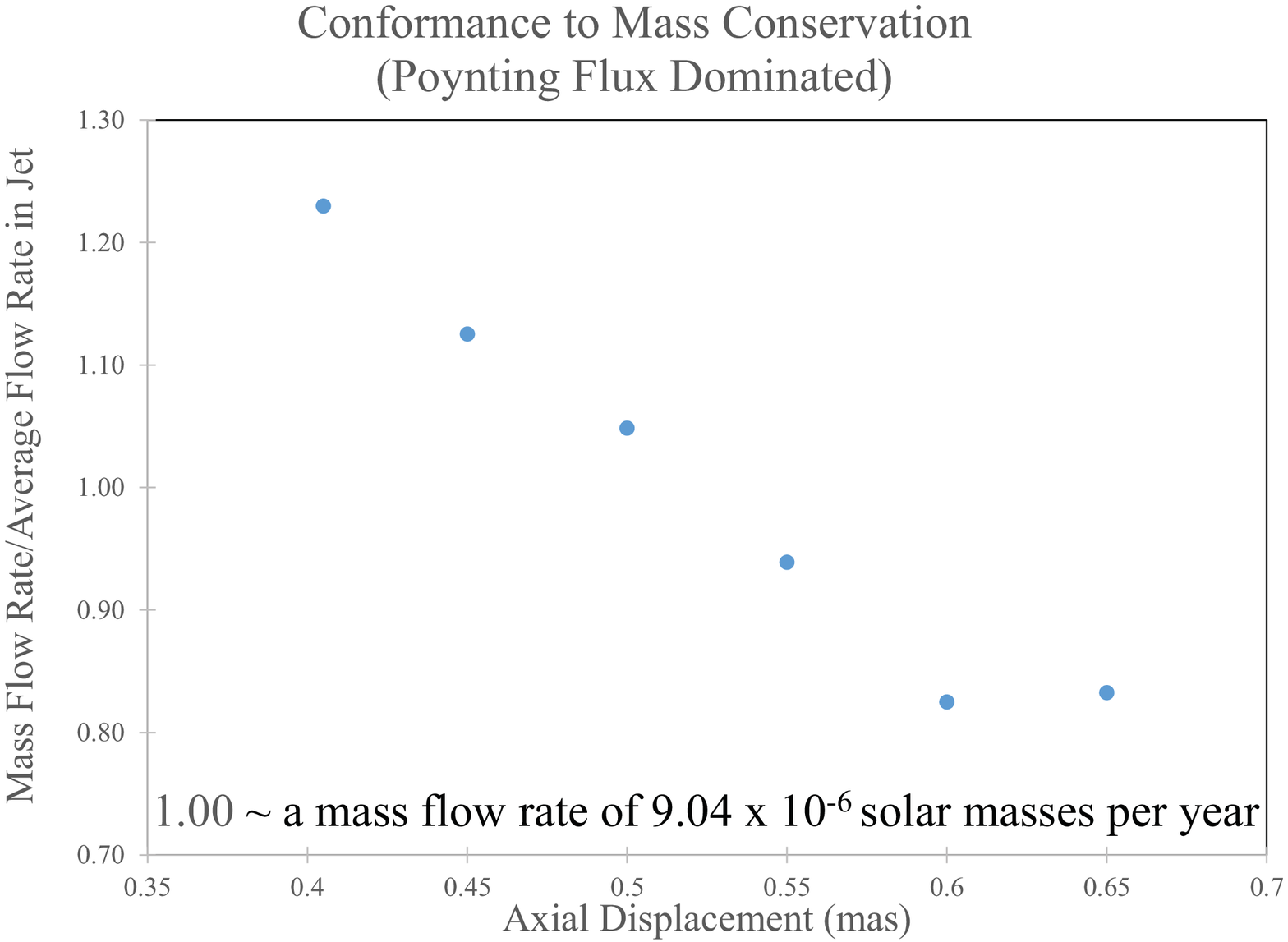}
\caption{The figure is used to determine how well the conservation laws are met by different hypotheses. The minimum energy solution violates both conservation laws and the Poynting jet dominated solution violates mass conservation.}
\end{center}
\end{figure*}

\section{Physical Realizations of the Jet Properties}
As a preliminary step, we review the physical expressions that are needed. For $z< 1$ mas, it was previously found that $\alpha \approx 0.6$ from 22 GHz to 43 GHz and $\alpha \approx 0.8$ from 43 GHz to 86 GHz, where the flux density, $S(\nu)\sim \nu^{-\alpha}$ \citep{had16}. Thus, $\alpha =0.7$ is adopted in the following.
The synchrotron emissivity is given in \citep{tuc75}:
\begin{eqnarray}
&& j_{\nu}(\nu_{e}) = 1.7 \times 10^{-21} [4 \pi N_{E}]a(n)B^{(1
+\alpha)}\left(\frac{4
\times 10^{6}}{\nu_{e}}\right)^{\alpha}\;,\\
&& a(n)=\frac{\left(2^{\frac{n-1}{2}}\sqrt{3}\right)
\Gamma\left(\frac{3n-1}{12}\right)\Gamma\left(\frac{3n+19}{12}\right)
\Gamma\left(\frac{n+5}{4}\right)}
       {8\sqrt\pi(n+1)\Gamma\left(\frac{n+7}{4}\right)} \;,
\end{eqnarray}
where the coefficient $a(n)$ separates the pure dependence on $n$ \citep{gin65}. A power law energy spectrum for the leptons with normalization $ N_{E}$ and number index, $n$ is assumed. The subscript ``e" stands for the emitted and ``o" for the observer's frames of reference. The total magnetic field is $B$. One can transform this to the observed flux density, $S(\nu_{\mathrm{o}})$, for an optically thin jet using the relativistic transformation relations from
\citet{lin85},
\begin{eqnarray}
 && S(\nu_{\mathrm{o}}) = \frac{\delta^{(2 + \alpha)}}{4\pi D^{2}}\int{j_{\nu}^{'}(\nu_{o}) d V{'}}\;,
\end{eqnarray}
where $D$ is the distance and in this expression, the
primed frame is the rest frame of the plasma. $\beta$ is the three-velocity of the moving plasma (from Table 1) and the Doppler factor, $\delta=1/[\gamma (1-\beta\cos{\theta})]$, with $\theta$ being the LOS, $\gamma = 1/\sqrt{1-\beta^{2}}$, and $\nu_{o}/\nu_{e} = \delta$.
\par $\mathcal{K}(\mathrm{protonic})$, $\mathcal{L}(\mathrm{protonic})$, $\mathcal{M}$ are the kinetic energy, angular momentum and mass flux of a protonic plasma, respectively:
\begin{equation}
 \mathcal{K}(\mathrm{protonic}) = \mathcal{M}c^{2}(\gamma-1)\;, \;\mathcal{L}(\mathrm{protonic}) = \mathcal{M}c\gamma\beta^{\phi}R_{\perp} \;, \;\mathcal{M}= Nm_{p}c\gamma \beta^{z}\;,
\end{equation}
where $m_{p}$ is the mass of the proton, $N$ is the proper number density, $\beta^{i}$ are components of the bulk velocity and $R_{\perp}$ is a radius in cylindrical coordinates. The lepto-magnetic energy, $E(\mathrm{lm})$, is the volume integral of the leptonic internal energy density, $U_{e}$, and the magnetic field energy density, $U_{B}$:
\begin{eqnarray}
 && E(\mathrm{lm}) = \int{(U_{B}+ U_{e})}\, dV = \int{\left[\frac{B^{2}}{8\pi}
+ \int_{E_{min}}^{E_{max}}(m_{e}c^{2})(N_{E}E^{-n + 1})\, d\,E \right]dV}, \, \nonumber \\
&& \mathcal{E}=\gamma\mathcal{M}c^{2}\left[\frac{4U}{3Nm_{e}c^{2}}-\frac{1}{3}-\gamma^{-1}\right]\;,
\end{eqnarray}
where $dV$ is the volume element and $U=U_{e}+U_{B}(\rm{turbulent \, fields})$. $\mathcal{E}$ is the free thermo-kinetic energy flux, a quantity that can be converted to and from Poynting flux \citep{mck12,rey20}.
\par In the electromagnetic sector, we consider the time stationary, axisymmetric approximation to the frozen-in conditions \citep{pun08},
\begin{equation}
E^{\perp} = -\beta_{F}B^{P}\;, \; \beta_{F}\equiv (\Omega_{F}R_{\perp})/c\; ,\quad B^{\phi}=\frac{\beta^{\phi}-\beta_{F}}{\beta^{P}}B^{P}\;,
\end{equation}
where ``P" indicates the poloidal direction of $B$ and ``$\perp$" is the orthogonal poloidal direction. $\Omega_{F}$ is the angular velocity of the magnetic field in the observer's frame (a constant).
The angular momentum, $S_{L}$, and the poloidal Poynting power, $S^{P}$, are \citep{pun08},
\begin{equation}
S_{L} = \frac{1}{4\pi}\int{B^{T}B^{P}dA_{\perp}}\;, S^{P} = \frac{1}{4\pi}\int{\Omega_{F}B^{T}B^{P}dA_{\perp}}\;,\quad \Phi \equiv \int{B^{P}dA_{\perp}}\;,
\end{equation}
where $\Phi$ is the conserved poloidal flux which defines $B^{P}$ (in any reference frame) in terms of the normal cross-sectional area element, $dA_{\perp}$. The identification of $B^{T}=R_{\perp}B^{\phi}$ is very useful because (due to mass conservation and $\Phi$ conservation) this is the conserved specific angular momentum flux along each field line in a very magnetically dominated perfect MHD wind \citep{pun08}.
The relativistic version, in Boyer-Lindquist coordinates (used throughout unless otherwise stated), (t, r, $\theta$, $\phi$) is \citep{pun08}:
\begin{equation}
B^{T}=[\sqrt{r^{2}-2Mr+a^{2}}\sin{\theta}]\sqrt{F^{r\,\theta}F_{r\, \theta}} \;, \;-M\leq a \leq M\;,
\end{equation}
where $a$ is the angular momentum per unit mass of the black hole and $F^{\mu\,\nu}$ is the Faraday tensor.
\par Complex dynamics likely exist within the walls, but the observations lack the resolution to motivate a particular detailed physical model. Thus, we approximate the jet by 6 (as in Table 1) 0.16 mas tall, overlapping, axisymmetric, thick-walled cylindrical volume elements, each with uniform plasma properties. Each of the 0.16 mas tall, modelled sections of the tube average jet properties emitted from the nucleus over $\sim 4-6 $ month intervals \citep{pun21}. First, consider a minimum energy solution (top panels of Figure 2) with $E(\mathrm{lm})$ minimized. We take $E_{\rm{min}}\approx \rm{m}_{\rm{e}}\rm{c}^2$ as there is no obvious reason why the particles would not extend to mildly relativistic energies. We also note that \citet{cel08} have argued that $E_{\mathrm{min}}=m_{e}c^2$ based on fits to blazar jet spectra in the soft X-ray band. We assume a turbulent magnetic field and a protonic medium. The power increases significantly and the mass flow rate, $\dot{\mathcal{M}}$, decreases significantly along the jet. Thus, the proposed plasma state does not satisfy energy or mass conservation and is disfavored. The kinetic power also increases along a positronic jet solution.
\par Next, consider a very magnetically dominated solution of almost pure Poynting flux. From Equations (7) and (8), $B \approx B^{\phi}$. In the bottom row of Figure 2, energy, but not mass, is conserved. $\dot{\mathcal{M}}$ decreases with a $\pm20\%$ range from the mean. The situation is the same for a leptonic plasma. Although not impossible, this disfavors this class of models that can transport $\sim 10^{43}$ergs/sec.
\par Figure 3 shows the details of a solution that satisfies energy, angular momentum, and mass conservation to within 4\%. The jet power is $Q=\int{\mathcal{K}(\mathrm{protonic})dA_{\perp}}+S^{P}=5.25 \pm 0.49\times 10^{41}$ ergs/sec in each hemisphere. The uncertainty propagates from that in columns (7) and (9) of Table 1 and Equations (2)-(8). Considering the crude approximation in Equation (1) to an actual jet, we note that $Q$ and $\dot{\mathcal{M}}$ vary approximately $\sim W^{\frac{1.7}{3.7}}$, so even a 25\% error in the $W$ estimate only changes $Q$ and $\dot{\mathcal{M}}$ by $\sim 10\%$. We note that there is no analogous leptonic jet solution. Due to using $\mathcal{E}$ instead of $\mathcal{K}(\mathrm{protonic})$, the only conservative solutions are very dominated by a nearly constant lepton energy flux ($S^{P}<< \int{\mathcal{E}\, dA}\approx Q$), the opposite of the bottom row of Figure 2. Without a significant $S^{P}$ to accelerate the plasma from the point of origin, these solution have no physical basis to occur. Thus, we consider the protonic solution more viable.

\begin{figure}[htp!]
\begin{center}
\includegraphics[width=85mm, angle=0]{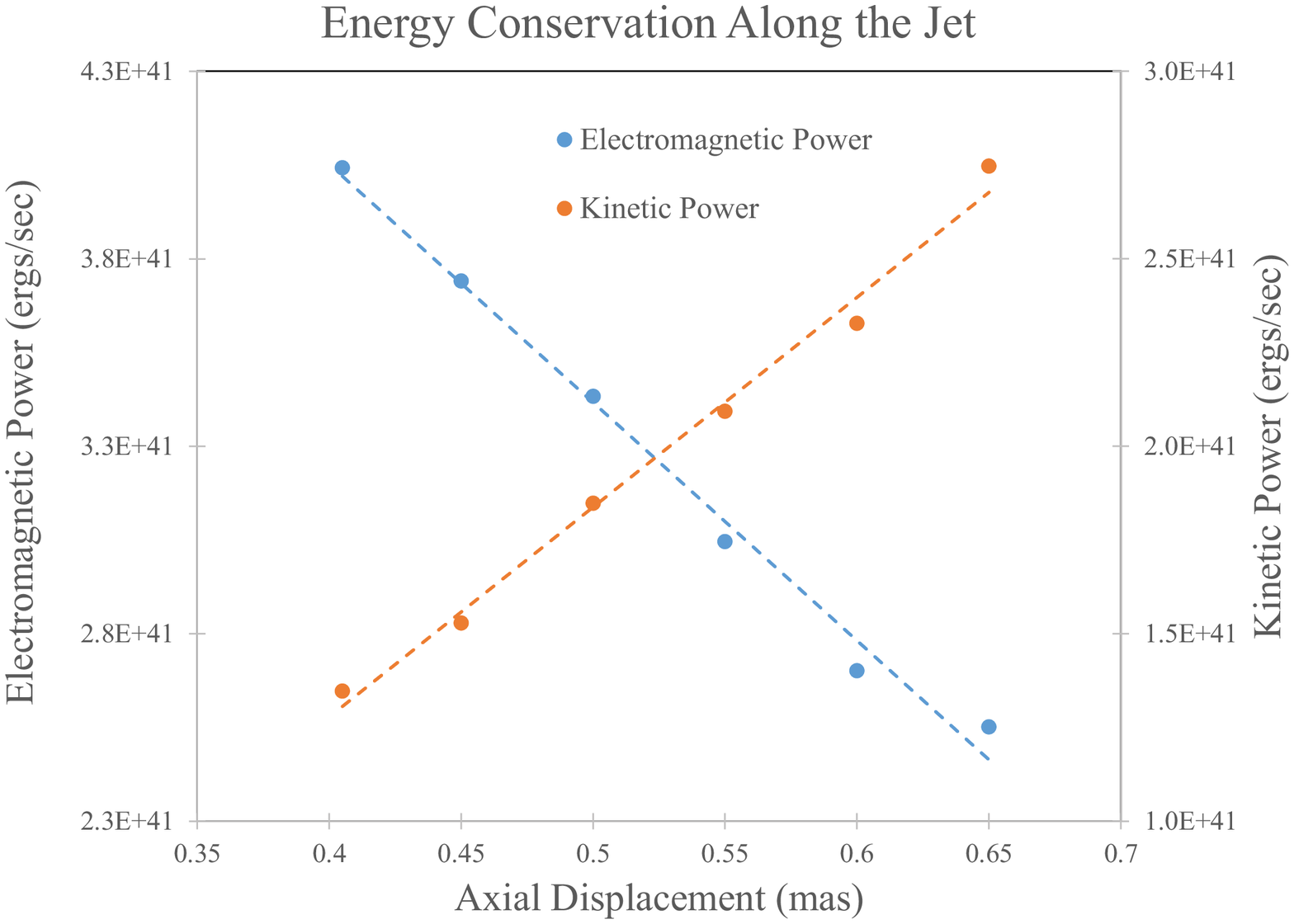}
\includegraphics[width=85mm, angle=0]{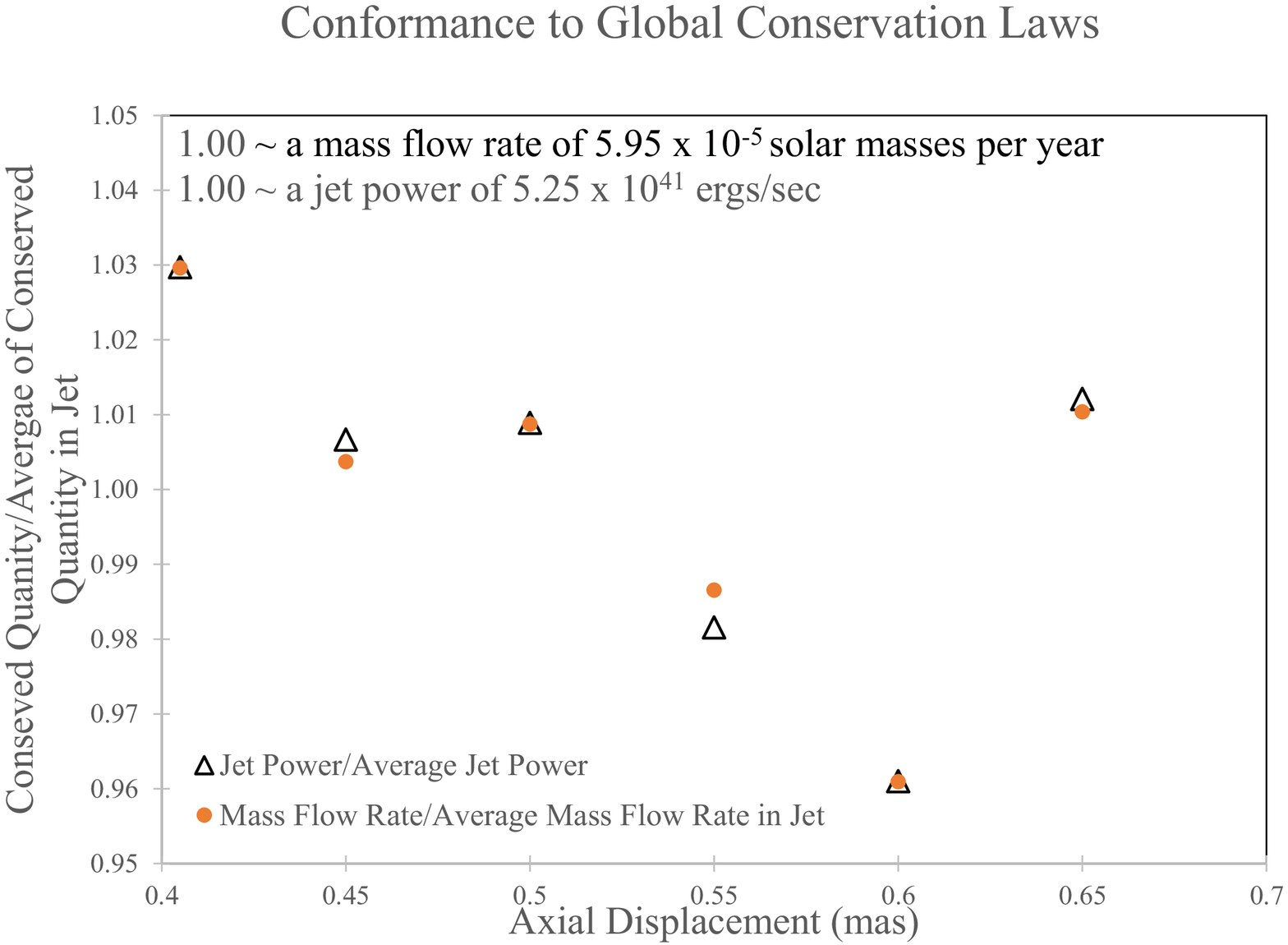}
\includegraphics[width=85mm, angle=0]{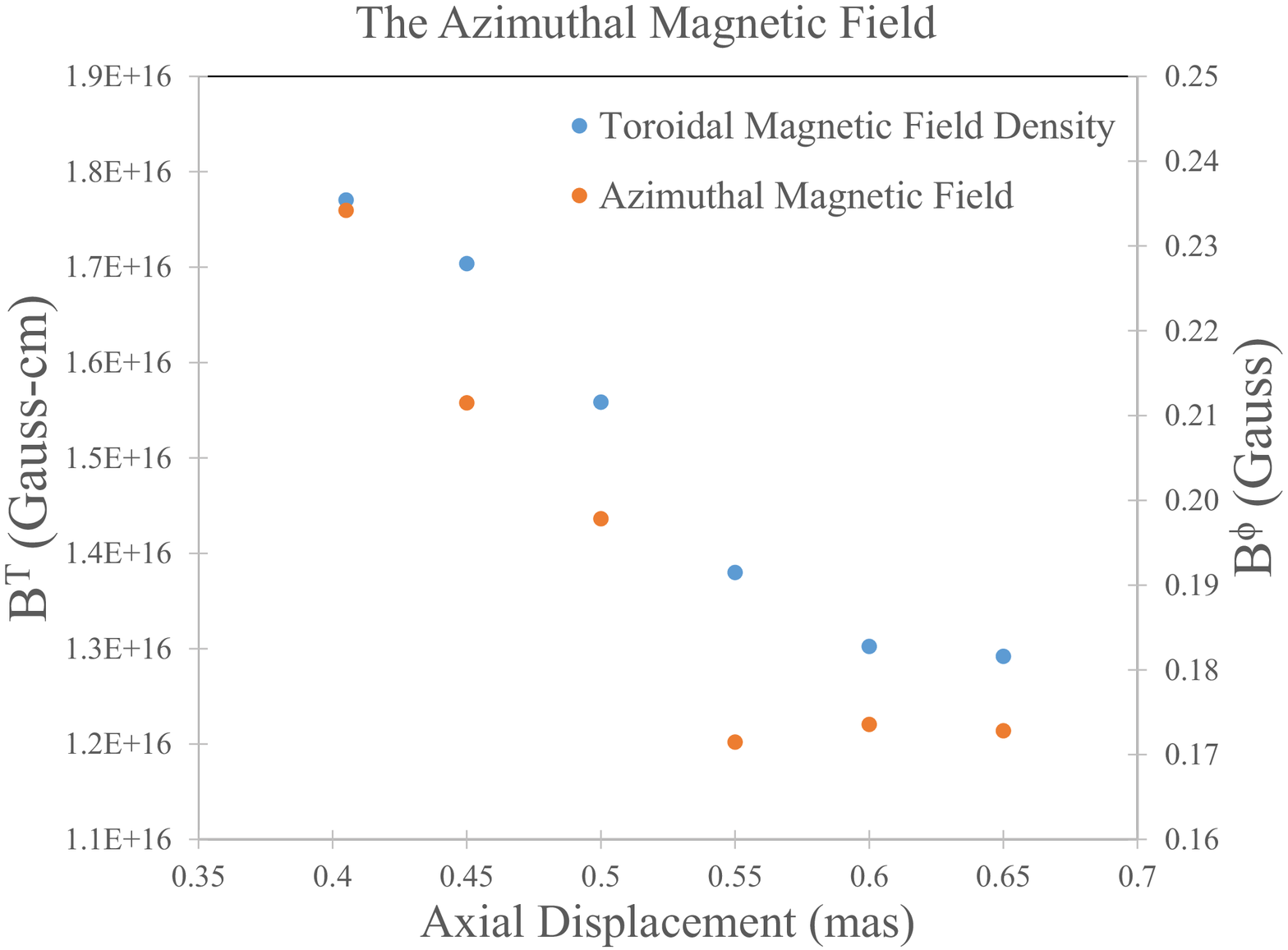}
\includegraphics[width=85mm, angle=0]{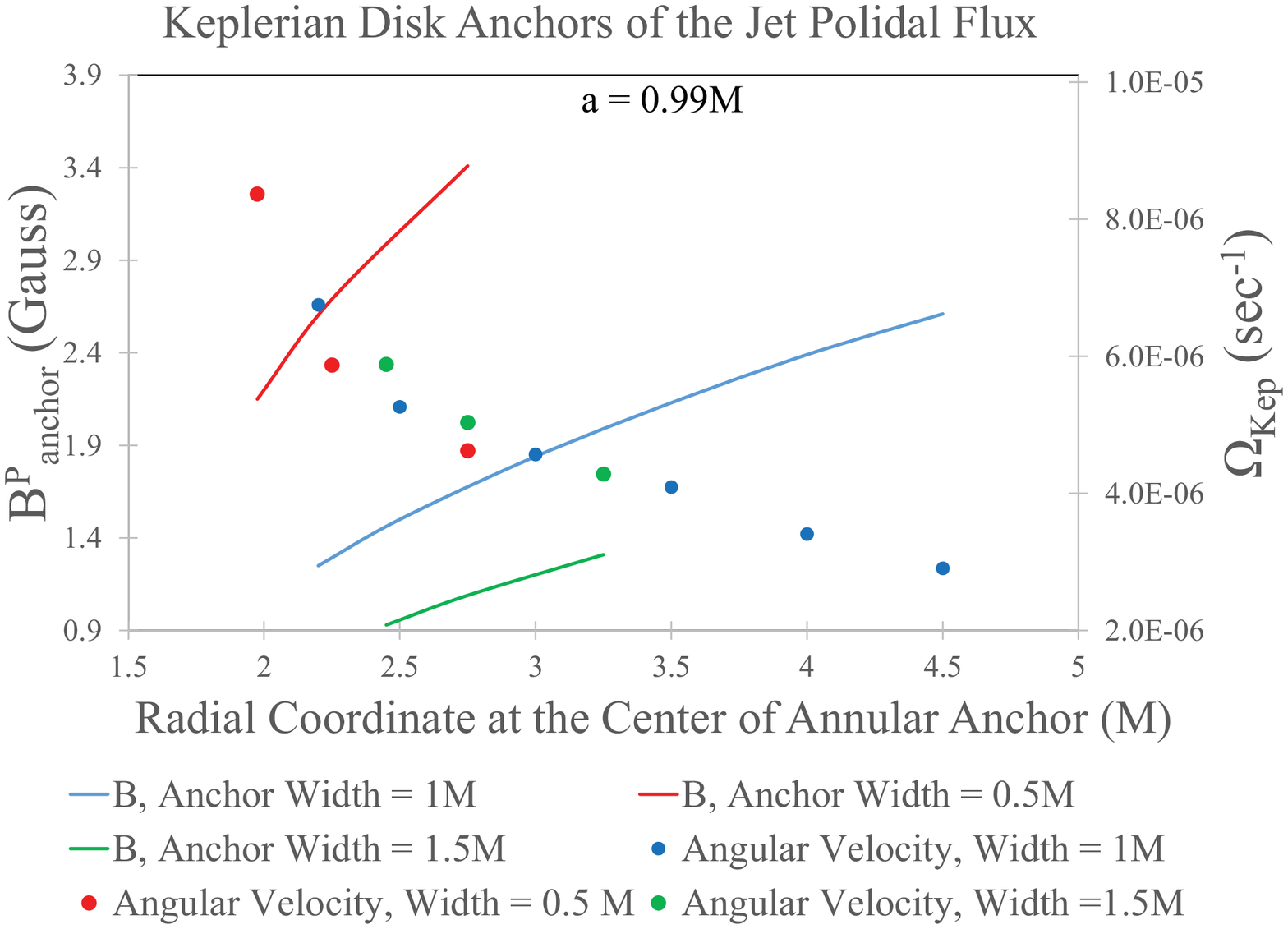}
\includegraphics[width=85mm, angle=0]{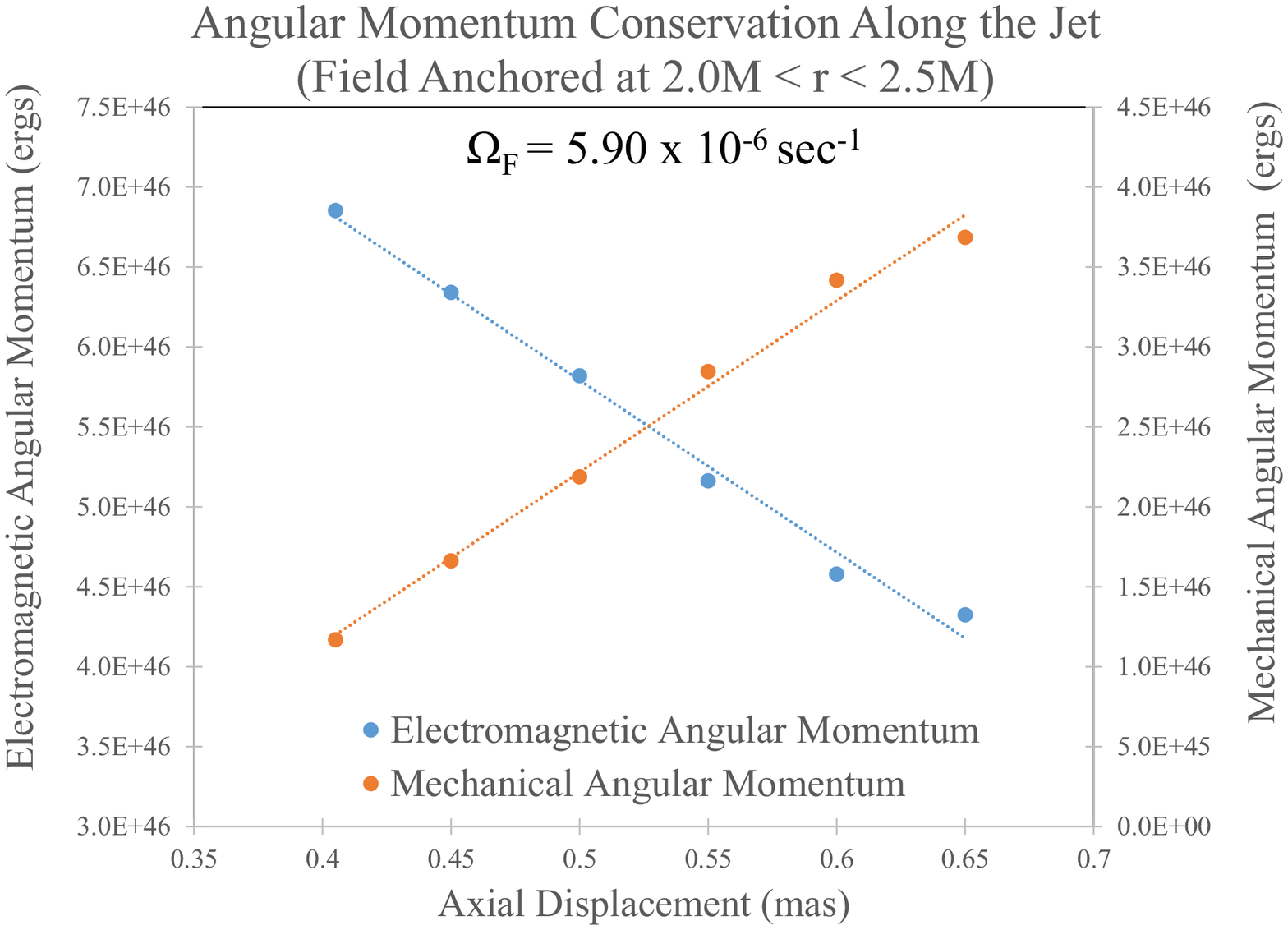}
\includegraphics[width=85mm, angle=0]{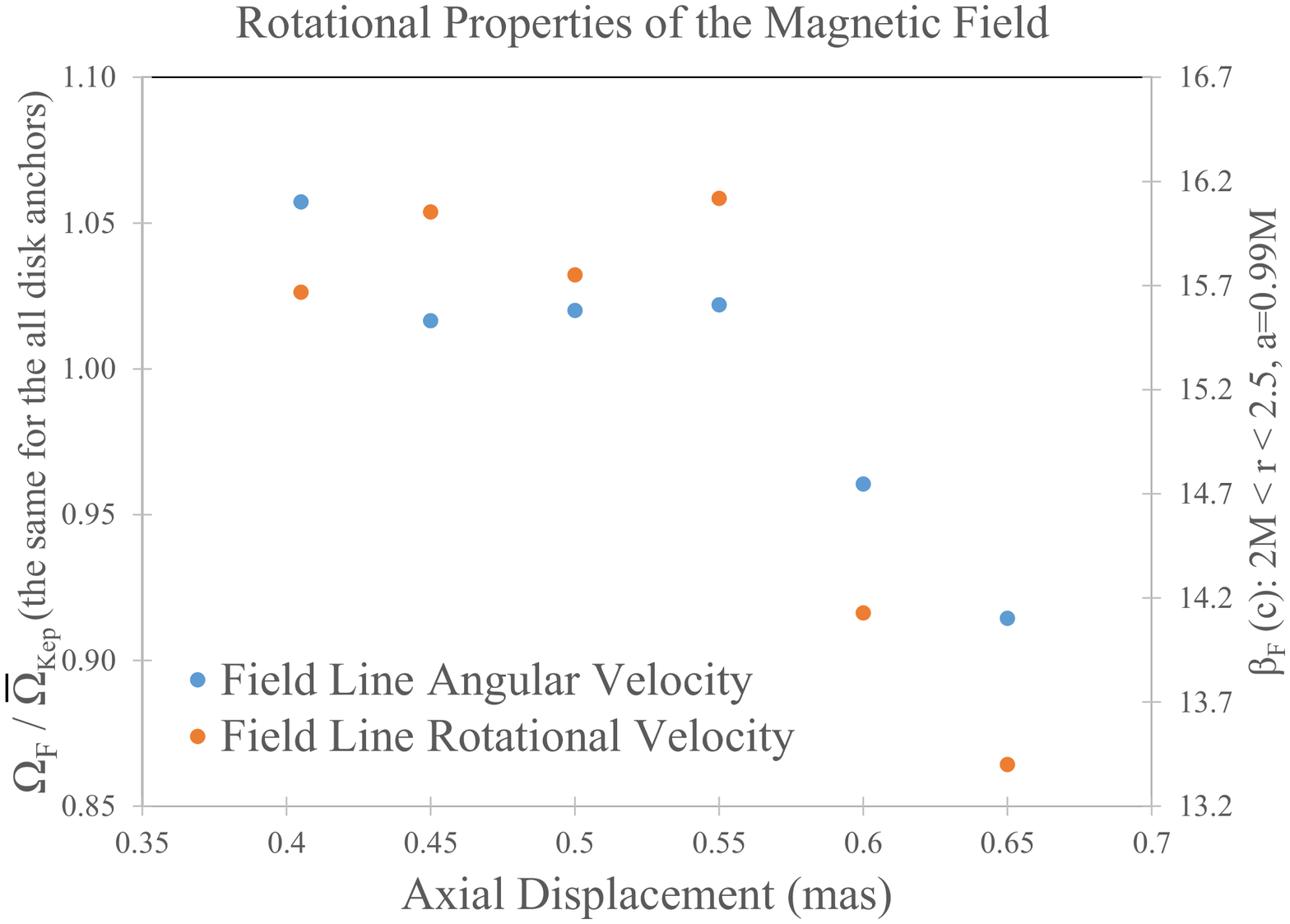}
\caption{A solution exists that satisfies all conservation laws (top row and bottom left panel). The top row and the middle left plot are determined without a knowledge of the anchor point. The middle right panel shows the anchor point condition on $B^{P}_{\rm{anchor}}$ for different Keplerian disk anchoring regions (``width" is the range of r coordinate). By choosing our fiducial region, $2.0M<r<2.5M$, as motivated in Section 5, we can complete the solution in the bottom row. The bottom right hand panel shows that $\Omega_{F}$ (the plot is the same independent of anchoring region) in our solution varies from $\overline{\Omega}_{\rm{Kep}}$ by $< 9\%$.}
\end{center}
\end{figure}

\section{The Source of the Solution that Obeys the Conservation Laws}
The largest reservoir of protons for the jet is the accretion disk. Thus, we consider the possibility that $\Phi$ is anchored to plasma that rotates, approximately, with the local Keplerian angular velocity \citep{lig75},
\begin{equation}
\Omega_{\rm{kep}}(r) = \frac{M^{0.5}}{r^{1.5}+aM^{0.5}}\;.
\end{equation}
The trend in the top left hand frame of Figure 3 indicates that the jet is very magnetically dominated just downstream of its source. We rewrite Equation (8) with the aid of the top left panel of Figure 3 as
\begin{equation}
S^{P}(\rm{from\, disk}) = \frac{1}{4\pi}B^{T}_{o}\int_{\rm{Anchor}}{\Omega_{F}B^{P}dA_{\perp}}\approx \frac{B^{T}_{o}B^{P}_{\rm{anchor}}}{4 \pi}\int_{\rm{Anchor}}{\Omega_{\rm{Kep}}dA_{\perp}}\approx 5.25 \times 10^{41} \rm{ergs/s}\;.
\end{equation}
$B^{T}_{o}$ is evaluated in the magnetically dominated wind just above the disk source (not in the disk source, where it is created). We pull $B^{T}_{o}$ out of the integral and evaluate it in the wind, then further approximate by treating $B^{P}_{\rm{anchor}}$ as a constant (uniform approximation). However, $\Omega_{\rm{Kep}}$ varies too rapidly near the black hole to be approximated as constant in the integrand.
Equation (8) and energy conservation imply
\begin{equation}
Q \approx \frac{1}{4\pi}B^{T}_{o}\Omega_{F}\Phi\; \rm{and}\; S^{P}(z)\approx \frac{1}{4\pi}B^{T}(z)\Omega_{F}\Phi\;,
\end{equation}
in the uniform tubular volume model. There can be only one value of $B^{T}_{o}$ which is approximated from the $B^{T}(z)$ values in the left middle of Figure 3 with Equation (12),
\begin{equation}
B^{T}_{o}(\rm{jet}) \equiv \frac{1}{6}\sum_{i=1}^{6}{B^{T}(z_{i})\frac{Q(z_{i})}{S^{P}(z_{i})}} = 2.43 \times 10^{16}\rm{G-cm}, z_{i} =0.405, 0.45,0.50,0.55,0.60,0.65\, \rm{mas}.
\end{equation}
Numerous values of $B^{P}_{\rm{anchor}}$ exist that solve Equations (7) and (11)-(13). Each choice corresponds to anchors with different $\Phi(\rm{anchor})$ and $\overline{\Omega}_{\rm{Kep}}$, that satisfy
\begin{equation}
\overline{\Omega}_{\rm{Kep}} \equiv \frac{\int_{\rm{Anchor}}{\Omega_{\rm{Kep}}dA_{\perp}}}{\int_{\rm{Anchor}}{dA_{\perp}}}=\frac{1}{6}\sum_{i=1}^{6}{\Omega_{F}(z_{i})}\equiv\overline{\Omega}_{F}\;,  z_{i} =0.405, 0.45,0.50,0.55,0.60,0.65\, \rm{mas}\;,
\end{equation}
where $\Omega_{F}(z)$ solves Equation (7), locally. Once the global $\overline{\Omega}_{F}$ is determined, angular momentum conservation follows from Equations (5), (7) and (8) in the bottom left panel of Figure 3. The global constants of the axisymmetric, time stationary MHD jet assumption, $\Phi$, energy, angular momentum and mass are conserved within $\pm 4\%$. Of all the conserved quantities, $\Omega_{F}(z)$ varies the most (the bottom right panel of Figure 3). Being the largest conservation law violation, its $\pm 9\%$ spread represents the error in our approximations and should taken as the uncertainty in all our calculated quantities due to the model. This figure also shows that $\beta_{F} >1$ (the analyzed region is beyond the light cylinder). $B^{P}_{\rm{anchor}}$ for a variety of different possible anchoring regions are indicated in the middle right panel of Figure 3. High spin, a = 0.99 M, was chosen so we can explore the region $r< 3$M and $\Omega_{\rm{kep}}$ in Equation (10) varies minimally with allowed spin values for $r>3$M.
\begin{figure}[htp!]
\begin{center}
\includegraphics[width=170mm, angle=0]{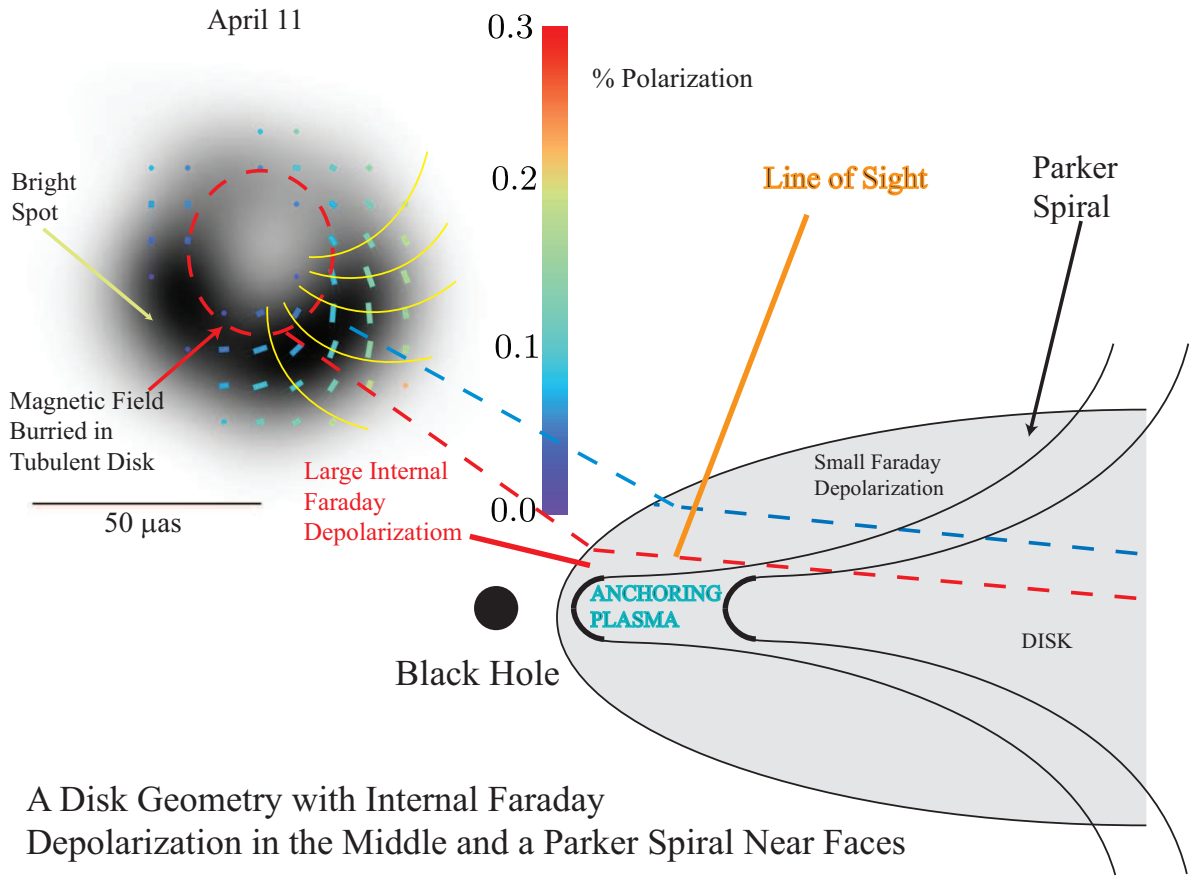}
\caption{The top left insert shows the polarization pattern (see text for details) from \citep{aki21}. We overlay some yellow Parker spiral type field lines to highlight a possible connection. The dashed red curve indicates a region of high Faraday depolarization. The cartoon in the lower right, shows the proposed connection between these features and the sources of the tubular jet found in the middle right panel of Figure 3.}
\end{center}
\end{figure}

\section{Is There a Connection to the EHT Polarization Pattern?} The middle, right panel of Figure 3 indicates candidate anchoring sites within the EHT annulus. In this section, we explore the compatibility of the tubular jet physical properties with the EHT annulus. We consider these circumstances:
\begin{enumerate}
\item The extension of the luminous tubular jet is $< 60$M (de-projected) above the EHT annulus \citep{had16,kim18,pun19}.
\item The bilateral jet transports $\approx 1.2\times10^{-4} M_{\odot}/\rm{yr}$ and the inner disk is a large reservoir of protons.
\item The magnetic field of a wind or jet is an Archimedean spiral, known as a Parker spiral based on his discovery in the solar wind \citep{par58}. Similarly, the polarization pattern of the EHT annulus is consistent with a mix of vertical, azimuthal, and radial ($B^{z}$, $B^{\phi}$ and $B^{r}$) fields \citep{nar21}. Resembling a face-on Parker spiral, the EHT polarization pattern can arise from $B^{\phi}$ and $B^{r}$ (LOS = $20^{\circ}$),  if $1.7B^{\phi}<B^{r}<3.7B^{\phi}$ \citep{nar21}.
\item A large $B^{r}$ (item 3) is consistent with the very large opening angle of the tubular jet at $0.07\rm{mas} < z < 0.25 \rm{mas}$, something that has been a challenge for published numerical simulations to replicate \citep{pun19}.
\end{enumerate}
\par Figure 4 shows the EHT polarization pattern from \citet{aki21} and a cartoon of the envisaged jet source that is compatible with the polarization properties. The length of the tick marks scale linearly with the polarized flux and the color gives us the $\%$ polarization from the color bar. The polarized flux is large (small) in the southwest (northeast) primarily due to Doppler boosting \citep{nar21}. The fractional polarization in the southwest and west decreases with radius. This is attributed to internal Faraday rotation by turbulent magnetic fields that results in significant internal Faraday depolarization and beam depolarization \citep{aki22,ric20}. We adopt the same depolarization prescription here. The dashed red curve marks a region of high depolarization.
\par We add some yellow magnetic field lines to the top left panel of Figure 4, similar to the line drawing of the Parker spiral in \citep{par58}. These are not calculated, only qualitative visual aids to highlight a possible spiral pattern in the polarization data. The spiral pattern would not be the spiral intrinsic to the wind base due to parallel transport of the polarization plane along the photon trajectories. There is a slight distortion, but a polarization pattern similar to that due to the intrinsic field results from parallel transport, see Figure 5 of \citep{nar21}. A vertical field is distorted so much by parallel transport that it can also make a similar polarization pattern \citep{nar21}. However, this latter scenario has no inferred direct connection to the adjacent tubular jet.
\par The bottom right hand frame is a cartoon of a field anchored in the leading edge of an accretion disk. Near the center of the disk, the turbulent magnetized plasma has sufficient column density along the line of sight to Faraday depolarize the synchrotron emission associated with organized field structure \citep{aki22}. The field anchored in the accretion flow exits the disk as a wind with a large radial field as required in magneto-centrifugally driven winds \citep{bla82}. The emission from the jet in the top layers of the disk (outer half of the EHT annulus) has high synchrotron polarization and modest Faraday depolarization. The bright spot indicated in the southeast of the annulus has no detected polarization and is evidence that polarization is being diluted by bright emission from a turbulent depolarized medium. The locally expected polarization should be $\gtrsim 70\%$, but values from 5\% -17\% are seen in Figure 4 \citep{aki22}. We therefore propose that $\sim 20-30\%$ of the emission is highly polarized jet emission that is diluted by the depolarized disk emission in the outer regions of the EHT annulus. It has been claimed that the EHT annulus can arise from a disk source 2M $<r<$ 4M \citep{aki20}. Based on the above discussion, we anchor the field at 2.0M $<r<$ 2.5M in our fiducial solution which solves the wind equations if $B^{P}_{\rm{anchor}}=2.7\rm{G}$ (see Figure 3). $B^{P}$ in the EHT annulus is unknown, but a constraint on the total field strength (including turbulent contributions) $\sim 1-30$\,G was argued in \citet{aki21} and a crude model indicated 4.9\,G in \citep{aki20}.

\section{Conclusion}
We have continued our study of an extraordinarily sensitive 43 GHz VLBA image by estimating the wall thickness of the tubular jet between 0.405 mas and 0.65 mas from intensity cross-cuts. Combining this analysis in Section 2 with the results of \citet{pun21}, we now have discrete estimates of the bulk velocity, acceleration, cross-sectional area and flux density along the tubular jet. In Section 3, we explore perfect MHD solutions that are defined by these physical properties. The primary result of this paper is a solution that satisfies, conservation of energy, angular momentum and mass. The most viable solution is protonic and magnetically dominated in which Poynting flux is converted to kinetic energy flux along the jet. The bilateral jet transports $\approx 1.2\times10^{-4} M_{\odot}/\rm{yr}$ (which bounds the accretion rate from below) and $Q\approx 1.1\times10^{42}$ erg/sec. The $\sim 10\%$ uncertainties in these quantities from the data and MHD modeling, determined in Section 3 and 4, are $\sim$ of fluctuations in the 22 and 43 GHz VLBI nuclear brightness during the epoch of jet ejection \citep{pun21}.
\par Plausible jet launching sites in the accretion disk are determined in Section 4. In Section 5, a speculative, circumstantial case is made for the jet being created in a Keplarian disk coincident with the source of the EHT annulus. It is based on the proximity of the jet to the disk, the EHT polarization pattern, the large supply of protons and $B^{z}\sim 1-3.5$G in our models, consistent with EHT collaboration estimates. The disk is not modelled due to poorly understood two-fluid (solar-type) plasma physics (even as basic as the electron temperature) \citep{aki22}. Therefore, no quantitative comparison to the parameters in Table 2 of \citet{aki22} can be made.

\begin{acknowledgments}
  We are thankful for the insightful comments of the referee. This work benefitted greatly from discussions with R. Craig Walker and Alan Marscher concerning the width of the intensity peaks and their relationship to wall thickness. R. Craig Walker also generously supplied the data for this paper. We also want to thank Krzysztof Nalewajko for pointing out that the bright spot in the EHT annulus is unpolarized. The Very Long Baseline Array (VLBA) is an instrument of the National Radio Astronomy Observatory. The National Radio Astronomy Observatory is a facility of the National Science Foundation operated by Associated Universities, Inc.

\end{acknowledgments}

\end{document}